\documentclass[aip,apl,reprint,groupedaddress,preprintnumber,showpacs]{revtex4-1}
\usepackage{graphicx}
\usepackage{color}
\usepackage{dcolumn}

\begin{document}



\title{Large linear magnetoresistance and magnetothermopower in layered SrZnSb$_2$}
\author{Kefeng Wang}
\affiliation{Condensed Matter Physics and Materials Science Department, Brookhaven National Laboratory, Upton New York 11973 USA}
\author{C. Petrovic}
\affiliation{Condensed Matter Physics and Materials Science Department, Brookhaven National Laboratory, Upton New York 11973 USA}

\date{\today}

\begin{abstract}
We report the large linear magnetoresistance ($\sim 300\%$ in 9 T field at 2 K) and magnetothermopower in layered SrZnSb$_2$ crystal with quasi-two-dimensional Sb layers. A crossover from the semiclassical parabolic field dependent magnetoresistance to linear field dependent magnetoresistance with increasing magnetic field is observed. The magnetoresistance behavior can be described very well by combining the semiclassical cyclotron contribution and the quantum limit magnetoresistance. Magnetic field also enhances the thermopower. Our results can be well understood by the magnetotransport of Dirac states in the bulk band structure.
\end{abstract}

\maketitle

The transport of low dimensional electronic systems is a central topic in contemporary condensed matter physics. Besides phenomena in some artificial two-dimensional (2D) structures such as quantum Hall effect in artificial electron gas and Dirac fermions in graphene,\cite{qh,graphene} bulk materials with layered structures can also host some low dimensional properties such as quasi-2D Dirac states. Recently, very large MR was observed in layered Bi-based compound SrMnBi$_2$ and CaMnBi$_2$ which is attributed to the anisotropic Dirac states.\cite{srmnbi21,srmnbi22,srmnbi23,camnbi21,camnbi22} In SrMnBi$_2$, MR approaches $\sim 200\%$ in 9 T field at 2 K.\cite{srmnbi21,srmnbi22} MR gives information about the characteristics of the Fermi surface and could point to candidate materials for magnetic memory or other spintronic devices. This makes SrMnBi$_2$ and the search for materials similar to SrMnBi$_2$ of considerable interest in both fundamental and application research.

SrMnBi$_2$ with 112-type structure contains MnBi$_4$ layers and 2D Bi square nets that alternate along the $c$-axis and are separated by Sr atoms.\cite{srmnbi21,srmnbi22,srmnbi23} Highly anisotropic Dirac states whose linear energy dispersion originates from the crossing of two Bi $6p_{x,y}$ bands in the double-sized Bi square net, were identified.\cite{srmnbi21} For Dirac fermions, the distance between the lowest and $1^{st}$ LLs in magnetic field, $\Delta_{LL}$, is very large and the quantum limit
where all carriers occupy only the lowest LL is easily realized in moderate fields.\cite{LL1,LL2} Consequently some quantum transport phenomena such as quantum Hall effect and large linear magnetoresistance (MR) could be observed by conventional experimental methods.\cite{qt1,qt2,qt3,qt4,qt5} Therefore it is of considerable interest to explore materials with bulk Dirac states.

Here we report detailed resistivity, Hall effect and thermopower measurements in SrZnSb$_2$ crystal with quasi-two-dimensional Sb layers. SrZnSb$_2$ is a bad metal with low carrier density and exhibits very large magnetoresistance ($\sim 300\%$ in 9 T field at 2 K) with a crossover from the semiclassical $MR \sim H^{2}$ below 2 T to $MR \sim H$ in higher fields. The magnetoresistance behavior can be described very well by combining the semiclassical cyclotron contribution and the quantum limit magnetoresistance. Magnetic field also has significant effect on thermopower. Our results indicate the possible existence of Dirac fermions in this material.


\begin{figure}[tbp]
\includegraphics[scale=0.4]{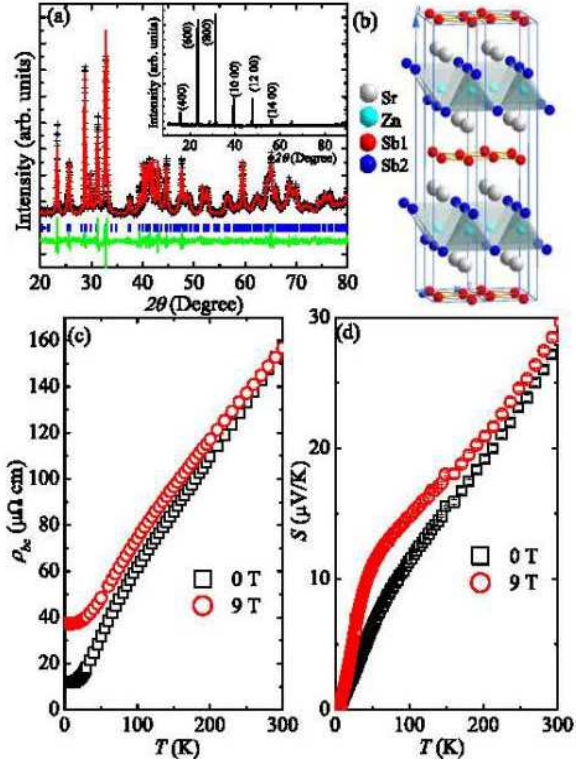}
\caption{(Color online) (a) Powder XRD patterns and structural refinement results. The data were shown by ($+$) , and the fit is given by the red solid line. The difference curve (the green solid line) is offset. Inset: The single crystal XRD pattern shows the basal plane of crystal is $bc$-plane. (b) The crystal structure of SrZnSb$_2$. Sb atoms in 2D layers (Sb1) are shows by red balls. Sr atoms are denoted by gray balls. Zn and Sb (Sb2) atoms in ZnSb$_4$ tetrahedrons (light blue polyhedrons in figure) are denoted by cyan and blue balls, respectively. Blue lines define the unit cell. (c) and (d) Temperature dependence of the in-plane resistivity $\protect\rho_{bc}(T)$ (c) and Seebck coefficient $S(T)$ (d)in $B=0$ T (squares) and $B=9$ T (circles) magnetic field respectively.}
\end{figure}

Single crystals of SrZnSb$_2$ were grown using a high-temperature self-flux method.\cite{flux} The resultant crystals are plate-like. X-ray diffraction (XRD) data were taken with Cu K$_{\alpha}$ ($\lambda=0.15418$ nm) radiation of Rigaku Miniflex powder diffractometer. Electrical transport measurements up to 9 T were conducted in Quantum Design PPMS-9 with conventional four-wire method. In the in-plane resistivity and Hall measurements, the current path was in the \textit{bc}-plane, whereas magnetic field was parallel to the \textit{a}-axis except in the angular dependent MR measurement. Seebeck coefficient was measured using steady state method and one-heater-two-thermometer setup with silver epoxy contact directly on the sample surface. The heat and electrical current were transported within the \textit{bc}-plane of the crystal, with magnetic field along the \textit{a}-axis and perpendicular to the heat/electrical current. The relative error in our measurement for both $\kappa$ and $S$ was below $5\%$ based on Ni standard measured under identical conditions.


Fig. 1(a) shows the powder XRD pattern of flux grown SrZnSb$_2$ crystals, which were fitted by RIETICA software.\cite{rietica} All reflections can be indexed in the Pnma space group. There are ZnSb$_4$ tetrahedral layers separated by the Sr atoms and 2D Sb blocking layers (Fig. 1(b)). Single crystal XRD pattern (inset of Fig. 1(a)) shows that the basal plane of a cleaved crystal is the crystallographic $bc$-plane. The in-plane resistivity $\rho_{bc}(T)$ shown in Fig. 1(c) exhibits a metallic behavior. An external magnetic field significantly enhances the resistivity by nearly three orders of magnitude. As the temperature is increased, MR is gradually suppressed and becomes rather small above $\sim 200$ K. Seebeck coefficient of SrZnSb$_2$ is positive in the whole temperature range indicating hole-type carriers (Fig. 1(d)).

SrZnSb$_2$ exhibits large magnetoresistance. The in-plane magnetoresistance MR=$(\rho_{bc}(B)-\rho_{bc}(0))/\rho_{bc}(0)$ reaches about $300\%$ in 9 T field at 2 K (Fig. 2(a)). The magnetoresistance is suppressed gradually with increase in temperature as shown in Fig. 2(a) but still is $\sim50\%$ at 100 K (Fig. 2(b)). More interesting, MR in SrZnSb$_2$ is linear in high field region. The inset in Fig. 2(a) shows the field derivative of the MR, $d$MR$/dB$, as a function of field $B$ at 40 K.  Below $\sim 2$ T, $d$MR$/dB$ is proportional to $B$, indicating the approximately quadratic field dependent MR ($\sim A_2B^2$). Above a characteristic field, $d$MR$/dB$ deviates from the semiclassical behavior and saturates to a much reduced slope indicating that the MR for $B>B^*$ is dominated by a linear field dependent term plus a very small quadratic term.

\begin{figure}[tbp]
\includegraphics[scale=1] {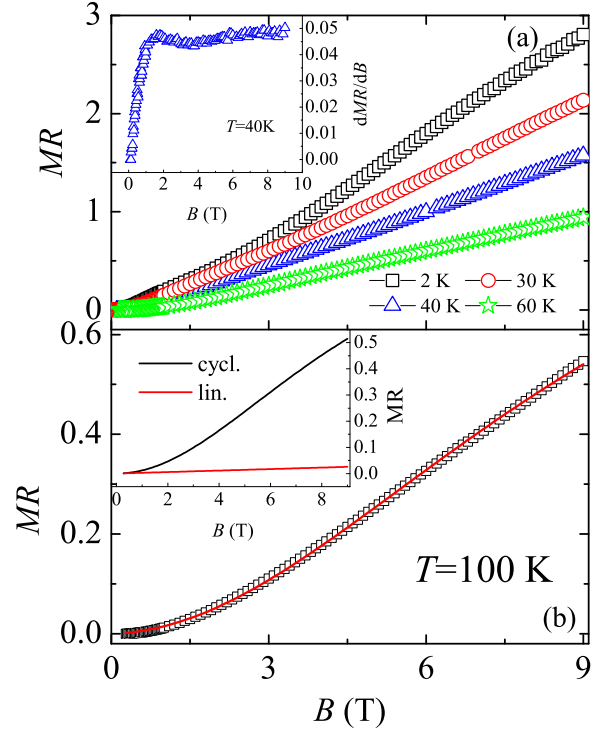}
\caption{(Color online) (a) The magnetic field ($B$) dependence of the in-plane magnetoresistance MR at different
temperatures. Inset: The field derivative of in-plane MR, $d$MR$/dB$, as a function of field (B) at $T=40 K$. (b) The fitting of the in-plane magnetoresistance at 100 K. The open squares are the experimental results and the solid line is the fitting result using Eqn.(1). In the inset the linear and the cyclotronic contributions are shown separately. }
\end{figure}

Fig. 3(a) shows the Hall resistivity $\rho_{xy}$ for SrZnSb$_2$ at different temperature. In the whole temperature range, $\rho_{xy}$ is positive which is consistent with Seebeck coefficient and indicates hole-type carriers in SrZnSb$_2$. The magnetic field dependent behavior in $\rho_{xy}$ is different from the classical Hall behavior. The positive $\rho_{xy}$ are not linear in field but quadratic. Fig. 3(b) shows the temperature dependencies of the Hall coefficient $R_H$ which is defined as $R_H=\rho_{xy}/B$ near $B=0$.\cite{hall1,mr} At low temperature, the apparent carrier density $n_{app}=\frac{R_H}{e}\sim2\times10^{18}$ cm$^{-3}$, which is smaller than conventional metals. Above $\sim 50$ K, $R_H$ (or the carrier density) shows a thermally activated behavior (Fig. 3(c)). These indicate that SrZnSb$_2$ is a bad metal with small carrier density and the Fermi level is very close to the valley in the density of states (DOS).\cite{hall1}

The large linear magnetoresistance and magnetothermopower effects in SrZnSb$_2$ are extraordinary and contradict the semiclassical transport theory. For conventional metals, the semiclassical transport gives $MR=\frac{\alpha B^2}{\beta+B^2}$.\cite{mr} This usually produces quadratic field-dependent MR in the low field range which would saturate in the high field, and the MR is always very small. If there are open orbits or Fermi surfaces, unsaturated MR with $B^2$-dependent would appear even in high field along the opened orbits while in other directions MR should still saturate. Consequently the linear field-dependent MR could be expected in polycrystal due to the average effect.\cite{mr} The large linear MR in SrZnSb$_2$ single crystal clearly deviates from above two types of behavior.

Large linear MR in magnetic field below 9 T was also observed in some other materials, such as Bi$_2$(Te/Se)$_3$, graphene, Ag$_{2-\delta}$Te, as well as iron-based superconductors BaFe$_2$As$_2$ and La(FeRu)AsO.\cite{qt1,qt2,qt3,qt4,qt5} All these materials were found to host Dirac fermions with linear energy dispersion, and the linear MR was ascribed to the quantum limit of Dirac fermions. In high enough field and the extreme quantum limit where carriers occupy only the lowest Landau level (LL), a large linear MR could be expected with $\rho=\frac{1}{2\pi}(\frac{e^2}{\epsilon_{\infty}\hbar v_F})^2\frac{N_i}{en^2}\mu_0H\ln(\epsilon_{\infty})$ where $v_F$ is the Fermi velocity, $N_i$ is the impurity concentration and $\epsilon_{\infty}$ is the high frequency dielectric constant.\cite{quantummr,quantumtransport} $\Delta_{LL}$ of Dirac fermions in magnetic field is very large and the quantum limit is easily realized in low field region.\cite{LL1,LL2}

\begin{figure}[tbp]
\includegraphics [scale=1]{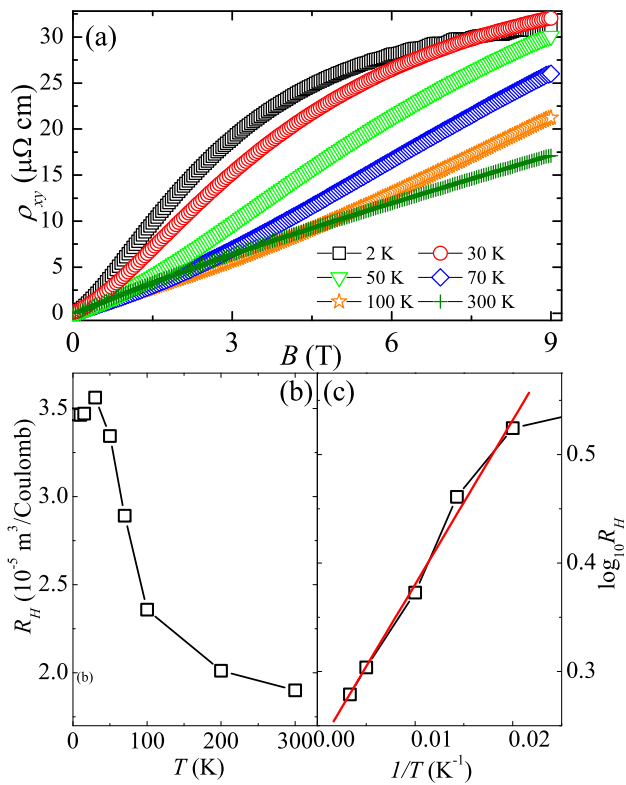}
\caption{(Color online) (a) The magnetic field ($B$) dependence of the Hall resistivity $\rho_{xy}$ at different temperatures. (b) Temperature dependence of the low field Hall coefficient $R_H$ for SrZnSb$_2$. (c) Arrhenius plot of $R_H(T)$ at high temperature range and the red line is the Arrhenius-law fitting.}
\end{figure}

Large linear MR was also observed recently in SrMnBi$_2$. In SrMnBi$_2$, highly anisotropic Dirac states were identified where linear energy dispersion originates from the crossing of two Bi $6p_{x,y}$ bands in the double-sized Bi square nets. The crystal structure of SrZnSb$_2$ features quasi-two-dimensional Sb layers similar to the double-sized Bi square nets in SrMnBi$_2$. It could be expected that the Sb layers in SrZnSb$_2$ can also host Dirac fermions. We note that the physical properties of SrZnSb$_2$ are very similar to SrMnBi$_2$ and CaMnBi$_2$. SrMnBi$_2$ and CaMnBi$_2$ are bad metals since the Fermi level is close to the valley in the density of states and close to the Dirac cone in the band structure because of the covalent nature of Bi 6p bonds in the 2D Bi layers.\cite{srmnbi21,srmnbi22} The magnitude of the resistivity in SrZnSb$_2$ is similar to that in (Sr/Ca)MnBi$_2$ and Hall resistivity reveals small carrier density. Because the Fermi level is close the Dirac-cone-like point in the band structure of (Sr/Ca)MnBi$_2$, the shift of the Fermi level by the magnetic field induces significant change in the carrier number and subsequently enhances significantly the absolute value of Seebeck coefficient. This is absent in conventional metals with very wide band crossing the Fermi level.\cite{thermopower} Similar effects was also observed in Ag$_{2-\delta}$Te.\cite{agte} In SrZnSb$_2$, the external magnetic field has small influence on the $S$ below $\sim 5$ K and above 250 K, but significantly enhances $S$ between 10 K and 200 K. This implies the existence of Dirac fermions in SrZnSb$_2$.

Taking the linear MR induced by quantum limit into account, MR in Dirac materials can be described by
\begin{eqnarray}
MR=\frac{\alpha B^2}{\beta+B^2}+\gamma|B|,
\end{eqnarray}
with $\alpha, \beta, \gamma$ as the fitting parameters.\cite{mr,quantummr,qt5} The first term of Eqn.1 is the cyclotron motion of carriers and the second term is the quantum limit MR. MR of SrZnSb$_2$ can be described very well by this mechanism. Fig. 2(b) shows a fitting example of MR at 100 K using above equation, and the inset in Fig. 2(b) gives the MR contribution from cyclotron motion (the first term in Eqn.1) and quantum linear MR (the second term in Eqn.1) at 100 K, respectively. This is consistent with the existence of Dirac fermions in SrZnSb$_2$. At low temperature, the quantum limit MR dominates. With increasing temperature, thermal fluctuation smears out the LL splitting, and consequently quantum limit contribution becomes smaller and total MR vanishes gradually.

In summary, we performed detailed resistivity, Hall effect and thermopower measurements in SrZnSb$_2$ crystal with quasi-two-dimensional Sb layers. SrZnSb$_2$ is a bad metal with low carrier density and exhibits very large magnetoresistance ($\sim 300\%$ in 9 T field at 2 K). A crossover from the semiclassical parabolic field dependent magnetoresistance in low field region to linear field dependent magnetoresistance in high field region is observed. The magnetoresistance behavior can be described very well by combining the semiclassical cyclotron contribution and the quantum limit magnetoresistance. Magnetic field also enhances significantly the thermopower. Our results indicate that the large linear MR arises from Dirac states in linear bands of the bulk band structure.

\begin{acknowledgments}
We than John Warren for help with SEM measurements. Work at Brookhaven is supported by the U.S. DOE under contract No. DE-AC02-98CH10886.
\end{acknowledgments}


\begin{thebibliography}{99}
\bibitem{qh}
D. Yoshioka, \textit{The Quantum Hall effect}, (Springer-Verlag, Berlin, 2002).
\bibitem{graphene}
A. H. Castro Neto, F. Guinea, N. M. R. Reres, K. S. Novoselov, and A. K. Geim, Rev. Mod. Phys. \textbf{81}, 109 (2009).
\bibitem{srmnbi21}
J. Park, G. Lee, F. Wolff-Fabris, Y. Y. Koh, M. J. Eom, Y. K. Kim, M. A. Farhan, Y. J. Jo, C. Kim, J. H. Shim, and J. S. Kim, Phys. Rev. Lett. {\bf 107}, 126402 (2011).
\bibitem{srmnbi22}
K. Wang, D. Graf, H. C. Lei, S. W. Tozer and C. Petrovic, Phys. Rev. B {\bf 84}, 220401 (2011).
\bibitem{srmnbi23}
J. K. Wang, L. L. Zhao, Q. Yin, G. Kotliar, M. S. Kim, M. C. Aronson, and E. Morosan, Phys. Rev. B {\bf 84}, 064428 (2011).
\bibitem{camnbi21}
K. Wang, D. Graf, L. Wang, H. C. Lei, S. W. Tozer and C. Petrovic, Phys. Rev. B {\bf 85}, 041101 (2012).
\bibitem{camnbi22}
J. B. He, D. M. Wang and G. F. Chen, Appl. Phys. Lett. {\bf 100}, 112405 (2012).
\bibitem{LL1}
Y. Zhang, Z. Jiang, Y.-W. Tan, H. L. Stormer, and P. Kim, Nature \textbf{438}, 201 (2005).
\bibitem{LL2}
D. Miller, K. Kubista, G. Rutter, M. Ruan, W. de Heer, P. First, and J. Stroscio, Science \textbf{324}, 924 (2009).
\bibitem{qt1}
D.-X. Qu, Y. S. Hor, J. Xiong, R. J. Cava, and N. P. Ong, Science \textbf{329}, 821 (2010).
\bibitem{qt2}
J. G. Analytis, R. D. McDonald, S. C. Riggs, J.-H. Chu, G. S. Boebinger, and I. R. Fisher, Nature Phys. \textbf{6}, 960 (2010).
\bibitem{qt3}
A. A. Taskin, Z. Ren, S. Sasaki, K. Segawa, and Y. Ando, Phys. Rev. Lett. {\bf 107}, 016801 (2011).
\bibitem{qt4}
R. Xu, A. Husmann, T. F. Rosenbaum, M.-L. Saboung, J. E. Enderby, and P. B. Littlewood, Nature {\bf 390}, 57 (1997).
\bibitem{qt5}
I. Pallecchi, F. Bernardini, M. Tropeano, A. Palenzona, A. Martinelli, C. Ferdeghini, M. Vignolo, S. Massidda, and M. Putti, Phys. Rev. B {\bf 84}, 134524 (2011).
\bibitem{flux}
P. C. Canfield and Z. Fisk, Phil. Mag. B {\bf 65}, 1117 (1992).
\bibitem{rietica} B. Hunter, "RIETICA - A Visual RIETVELD Progarm,"
International Union of Crystallography Commission on Powder Diffractin
Newsletter No. 20 (Summer), 1998.
\bibitem{hall1}
Z. Ren, A. A. Taskin, S. Sasaki, K. Segawa, and Y. Ando, Phys. Rev. B {\bf 85}, 155301 (2012).
\bibitem{mr}
A. B. Pippard, \textit{Magnetoresistance in Metals} (Cambridge University, Cambridge, 1989).
\bibitem{quantummr}
A. A. Abrikosov, Phys. Rev. B \textbf{58}, 2788 (1998).
\bibitem{quantumtransport}
A. A. Abrikosov, Europhys. Lett. \textbf{49}, 789 (2000).
\bibitem{thermopower}
K. Wang, L. Wang and C. Petrovic, Appl. Phys. Lett. {\bf 100}, 112111 (2012).
\bibitem{agte}
Y. Sun, M. B. Salamon, M. Lee, and T. F. Rosenbaum, Appl. Phys. Lett. {\bf 82}, 1440 (2003).
\end{thebibliography}

\end{document}